\documentclass[preprint]{emulateapj}

\shorttitle{Discovery of a Very Highly Extinguished Supernova}
\shortauthors{Kankare et al.}

\def\kms{\ifmmode{\rm km\,s^{-1}}\else\hbox{$\rm km\,s^{-1}$}\fi}

\begin{document}


\title{Discovery of a very highly extinguished supernova in a luminous infrared galaxy}


\author{E. Kankare\altaffilmark{1}, S. Mattila\altaffilmark{1}, S. Ryder\altaffilmark{2}, 
M.-A. P\'erez-Torres\altaffilmark{3}, A. Alberdi\altaffilmark{3}, C. Romero-Canizales\altaffilmark{3}, T. D{\'{\i}}az-Santos\altaffilmark{4}, P. V\"ais\"anen\altaffilmark{5}, A. Efstathiou\altaffilmark{6}, A. Alonso-Herrero\altaffilmark{4}, L. Colina\altaffilmark{4}, J. Kotilainen\altaffilmark{1}}
\altaffiltext{1}{Tuorla Observatory, Department of Physics and Astronomy, University of Turku, V\"ais\"al\"antie 20, FI-21500 Piikki\"o, Finland; eskank@utu.fi}
\altaffiltext{2}{Anglo-Australian Observatory, PO Box 296, Epping, NSW 1710, Australia.}
\altaffiltext{3}{Instituto de Astrofísica de Andalucia, IAA-CSIC, Apartado 3004, 18080 Granada, Spain}
\altaffiltext{4}{Instituto de Estructura de la Materia IEM-CSIC, 28006 E-Madrid, Spain}
\altaffiltext{5}{South African Astronomical Observatory, PO Box 9, Observatory 7935, South Africa.}
\altaffiltext{6}{School of Sciences, European University Cyprus, Diogenes street, Engomi, 1516 Nicosia Cyprus.}




\begin{abstract}
We report the discovery of a confirmed supernova (SN) and a supernova-candidate in near-infrared images from the ALTAIR/NIRI adaptive optics system on the Gemini-North Telescope and NICMOS on the \textit{Hubble Space Telescope}. The Gemini images were obtained as part of a near-infrared \textit{K}-band search for highly-obscured SNe in the nuclear regions of luminous infrared galaxies. \object{SN 2008cs} apparent in the Gemini images is the first SN discovered using laser guide star adaptive optics. It is located at 1500 pc projected distance from the nucleus of the luminous infrared galaxy \object{IRAS 17138-1017}. The SN luminosity, \textit{JHK} colors and light curve are consistent with a core-collapse event suffering from a very high host galaxy extinction of $15.7 \pm 0.8$ magnitudes in \textit{V}-band which is to our knowledge the highest yet measured for a SN. The core-collapse nature of SN 2008cs is confirmed by its radio detection at 22.4 GHz using our Very Large Array observations 28 days after the SN discovery, indicating a prominent interaction of the SN ejecta with the circumstellar medium. An unconfirmed SN apparent in the NICMOS images from 2004 is located in the same galaxy at 660 pc projected distance from the nucleus and has a lower extinction.
\end{abstract}

\section{Introduction}
A large fraction of the massive star formation at high-$z$, e.g., $~70 \%$ at $z = 1$ (Le Floc'h et al. 2005), took place in luminous (L$_{{\rm IR}}$ $>$ 10$^{11}$ L$_{\odot}$) and ultraluminous (L$_{{\rm IR}}$ $>$ 10$^{12}$ L$_{\odot}$) infrared (IR) galaxies (LIRGs and ULIRGs, respectively), and their high star formation rates (SFRs) can be expected to result in core-collapse supernova (CCSN) rates a couple of orders of magnitude higher than in ordinary field galaxies. Most of the SNe occurring in LIRGs and ULIRGs are likely to be obscured by large amounts of dust in the nuclear starburst environment; however, in the near-IR the extinction is strongly reduced. Because of the likely concentration of most of the SNe within the innermost nuclear regions a high spatial resolution is crucial for their detection. This can be achieved with space-based imaging (e.g., Cresci et al. 2007, Colina et al. 2007), imaging from Antarctica (see Burton et al. 2005) or ground-based adaptive optics (AO) imaging providing resolutions 5 times better than under typical natural seeing conditions. The potential of the current 8m-class telescopes equipped with AO was recently demonstrated by the discovery of SN 2004ip in the LIRG nuclear regions using the NACO AO system with a natural guide star on the Very Large Telescope (Mattila et al. 2007). 

In this letter we report a discovery of two SNe in \object{IRAS 17138-1017} which is a LIRG (L$_{IR} = 2.4 \times 10^{11}$ L$_{\odot}$, Sanders et al. 2003) at a redshift of $z$ = 0.0173 (Shier \& Fischer 1998) corresponding to a distance of 75 Mpc (H$_{0}$ = 70 km~s$^{-1}$Mpc$^{-1}$, $\Omega_{\Lambda}$=0.7 and $\Omega_{\rm M}$=0.3). The SNe were detected as a result of our \textit{K}-band search for highly-obscured CCSNe in a sample of nearby LIRGs using the Gemini-North Telescope. Even though \object{IRAS 17138-1017} has a high expected SN rate, only one SN (the Type Ia \object{SN 2002bw}, with a likely low extinction; Li 2002, Matheson et al. 2002) has been detected in this galaxy before.

\section{Observations and results}

\subsection{IR observations}
\object{IRAS 17138-1017} was observed with \textit{HST}/NICMOS as part of a program to observe a complete sample of LIRGs (Alonso-Herrero et al. 2006) with the NIC2 camera (0.076\arcsec\,pixel$^{-1}$, FWHM $\sim$ 0.1-0.2\arcsec) on 2004 September 23.0 UT (see Table~1). Observations were conducted with the F110W and F160W broadband filters and with the F187N narrowband filter. Details of the data reduction and photometric calibration of the NICMOS images can be found in Alonso-Herrero et al. (2006). The LIRG was next observed with the Gemini-North Telescope\footnote{Based on observations obtained at the Gemini Observatory, which is operated by the Association of Universities for Research in Astronomy, Inc., under a cooperative agreement with the NSF on behalf of the Gemini partnership: the National Science Foundation (USA), the Science and Technology Facilities Council (UK), the National Research Council (Canada), CONICYT (Chile), the Australian Research Council (Australia), Ministério da Ciência e Tecnologia (Brazil) and SECYT (Argentina).} in the \textit{K}-band using the Near-InfraRed Imager (NIRI) with the ALTAIR laser guide star AO system (0.022\arcsec\,pixel$^{-1}$, FWHM $\sim$ 0.1\arcsec) on 2008 April 21.6 UT (program GN-2008A-Q-38, PI: S. Ryder). The NIRI images were reduced using the NIRI package in {\sc iraf}. The target frames were de-dithered and median combined in {\sc iraf} using a bright field star as a reference point. Comparison between the \textit{HST} and Gemini observations revealed two new point sources, one at each epoch. The object apparent only in the \textit{HST} images is located at 1.9\arcsec\ (or 660 pc) and the one seen only in the Gemini images at 4.2\arcsec\ (or 1500 pc) projected distance from the near-IR nucleus of \object{IRAS 17138-1017}. Both objects were reported to the Central Bureau for Astronomical Telegrams (CBAT) immediately after discovery and the more recent one was later assigned the designation \object{SN 2008cs} (Kankare et al. 2008,  P\'erez-Torres et al. 2008). Follow-up observations of \object{SN 2008cs} were obtained on May 20.5 UT with NIRI in the \textit{J}, \textit{H}, and \textit{K} bands without AO (0.117\arcsec\,pixel$^{-1}$, FWHM $\sim$ 0.5\arcsec), and with ALTAIR/NIRI on May 27.5, June 25.5, and September 20.2 - 21.2 UT (see Fig.~1) using the AO. Follow-up observations were also obtained using the Nordic Optical Telescope (NOT) with the NOTCam instrument (0.234\arcsec\,pixel$^{-1}$, FWHM $\sim$ 0.7\arcsec) in the \textit{J}, \textit{H}, and \textit{K}$_{\rm s}$ bands on June 18.1 UT (see Table~1).

We used 35 stars from the 2MASS catalog to derive an accurate World Coordinate System (WCS) for an \textit{HST}/ACS \textit{I}-band image of \object{IRAS 17138-1017} obtained from the \textit{HST} Science Archive. The Gemini/NIRI \textit{K}-band (AO) image from June 25 was aligned to the ACS image using seven pointlike sources common between the frames and a simple geometric transformation including shifts in x and y, and the same scale factor and rotation for both x and y. This yielded $17^{\mathrm{h}}16^{\mathrm{m}}35.855^{\mathrm{s}}$ and $-10^{\circ}20\arcmin42.98\arcsec$ for \object{SN 2008cs} with 0.09\arcsec\ and 0.07\arcsec\ uncertainty in R.A. and Decl. respectively. The \textit{HST}/NICMOS F160W image was aligned to the ACS image using 6 point sources and this yielded R.A. $= 17^{\mathrm{h}}16^{\mathrm{m}}35.900^{\mathrm{s}}$ and Decl. $= -10^{\circ}20\arcmin37.85\arcsec$ for the 2004 event.

\begin{figure*}[t]
\epsscale{1.00}
\plotone{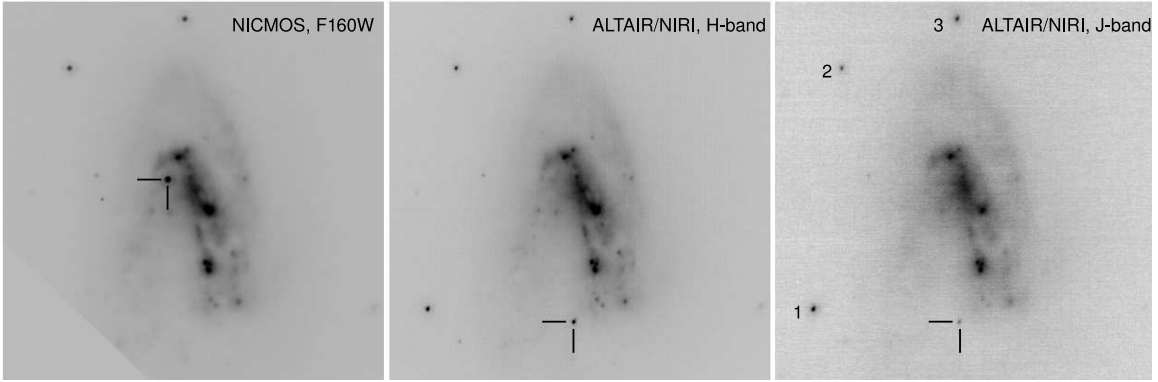}
\caption{14\arcsec$\times$14\arcsec subsections of IRAS 17138-1017 images taken with the \textit{HST}/NICMOS F160W filter on September 23, 2004, and with Gemini ALTAIR/NIRI AO in \textit{H}-band and \textit{J}-band on June 25, 2008. The ``historical'' SN from 2004 and SN 2008cs are indicated in the NICMOS and Gemini images, respectively. The effects of high extinction in the SN 2008cs line-of-sight can also be seen by comparing the brightness of the SN to the field stars in the \textit{H}- and \textit{J}-bands. The three field stars indicated in the right-hand panel were used for photometric calibration. North is up and East is to the left in all the images.}
\end{figure*}

\subsection{SN 2008cs photometry and extinction}
The photometry of \object{SN 2008cs} was measured from the images using the {\sc SNOOPY}\footnote{{\sc SNOOPY}, originally presented in Patat (1996), has been implemented in {\sc iraf} by E. Cappellaro. The package is based on {\sc daophot}, but optimised for SN magnitude measurements.} package in {\sc iraf}. Suitable field stars were selected to fit a point-spread function (PSF) for each image. A polynomial surface was fitted to a background region around the object position (excluding the innermost region around the object) and after removing the local background the PSF was fitted to the object and a residual image was produced after subtracting the PSF. Error estimates for the measurements were obtained by simulating and PSF fitting nine artificial point sources around the SN in the residual images. For the non-AO Gemini/NIRI and NOT images the photometry was calibrated using six 2MASS stars within the image field of view (FOV). The AO Gemini/NIRI images with small FOV were photometrically calibrated to the non-AO images using PSF photometry of three field stars available in both the AO and non-AO images (see Fig.~1). Assuming that these field stars have a similar AO PSF as the SN, this procedure also corrects for the flux lost in the PSF halo around the diffraction limited core in the AO images. Finally, all the magnitudes were aperture corrected using three bright isolated stars in the non-AO images to an aperture of 3\arcsec\ in radius. For the complete list of apparent SN magnitudes see Table~1.

\begin{table}[t]
\caption{Log of the observations and supernova photometry.}
\centering
\begin{tabular}{ccccc}
UT Date & Instrument \& Filter & Exp. time & Mag    \\
\tableline\tableline
2004 & & & SN candidate \\
\tableline
Sept. 23.0 & NICMOS, F110W & 3 $\times$ 80 s & $18.28 \pm 0.06$\\
Sept. 23.0 & NICMOS, F160W & 3 $\times$ 80 s & $17.00 \pm 0.05$\\
Sept. 23.1 & NICMOS, F187N & 3 $\times$ 288 s & $16.58 \pm 0.06$\\
\tableline\tableline
2008 & & & SN 2008cs \\
\tableline
April 21.6 & ALTAIR/NIRI, \textit{K} & 8 $\times$ 30 s & $16.65 \pm 0.06$\\
May 20.5 & NIRI, \textit{K} & 9 $\times$ 30 s & $16.22 \pm 0.37$\\
May 20.5 & NIRI, \textit{H} & 7 $\times$ 30 s & $17.40 \pm 0.46$\\
May 20.5 & NIRI, \textit{J} & 9 $\times$ 30 s & $19.43 \pm 0.50$\\
May 27.5 & ALTAIR/NIRI, \textit{K} & 8 $\times$ 30 s & $15.77 \pm 0.06$\\
May 27.5 & ALTAIR/NIRI, \textit{H} & 8 $\times$ 30 s & $16.97 \pm 0.06$\\
June 18.1 & NOTCam, \textit{K}$_{\rm s}$  & 18 $\times$ 60 s & $15.99 \pm 0.61$\\
June 18.1 & NOTCam, \textit{H} & 18 $\times$ 60 s & $16.92 \pm 0.35$\\
June 18.1 & NOTCam, \textit{J} & 18 $\times$ 60 s & $19.58 \pm 0.65$\\
June 25.5 & ALTAIR/NIRI, \textit{K} & 8 $\times$ 30 s & $15.82 \pm 0.10$\\
June 25.5 & ALTAIR/NIRI, \textit{H} & 8 $\times$ 30 s & $17.17 \pm 0.05$\\
June 25.5 & ALTAIR/NIRI, \textit{J} & 8 $\times$ 30 s & $19.02 \pm 0.14$\\
Sept. 20.2 & ALTAIR/NIRI, \textit{K} & 8 $\times$ 30 s & $17.30 \pm 0.09$\\
Sept. 20.2 & ALTAIR/NIRI, \textit{H} & 8 $\times$ 30 s & $19.11 \pm 0.09$\\
Sept. 21.2 & ALTAIR/NIRI, \textit{J} & 8 $\times$ 30 s & $>21.4$\\
\tableline
\end{tabular}
\end{table}

Assuming a distance of 75 Mpc and a Galactic extinction of A$_{\rm V}$ = 2.29 (Schlegel et al. 1998, but see also Arce \& Goodman 1999) the absolute magnitude of the SN at the time of the second AO observation (when the SN is closest to the peak) becomes M$_{\rm H}$ = $-17.81$ and M$_{\rm K}$ = $-18.87$ (before correcting for the host galaxy extinction). We compared these with the absolute peak magnitudes of typical Type II-P SNe 1999em and 2003hn (M$_{\rm H} \approx -17.8$ and M$_{\rm K} \approx -18.1$; Krisciunas et al. 2008) and the ``ordinary'' (M$_{\rm H}$ = $-18.58$ and M$_{\rm K}$ = $-18.62$) and ``slowly-declining'' (M$_{\rm H}$ = $-19.95$ and M$_{\rm K}$ = $-20.02$) CCSN template light curves from Mattila \& Meikle (2001). The observed SN color and peak absolute magnitude can only be accounted for by the ``slowly-declining'' type, and even then only with a substantial host galaxy extinction. Adopting these \textit{JHK} templates and the extinction law of Rieke \& Lebofsky (1985) for which A$_{\rm K}$ = 0.112 $\times$ A$_{\rm V}$ we obtain a host galaxy extinction towards \object{SN 2008cs} of $1.67 - 1.85$ mag in \textit{K}. This corresponds to $14.9 - 16.5$ mag in \textit{V} (A$_{\rm V} = 17.2 - 18.8$ total line of sight extinction) with the intrinsic absolute peak magnitude M$_{\rm K}$ = $-20.9 \pm 0.1$ (statistical error only) occurring around $49 - 55$ days after the discovery (see Fig.~2 for an example fit). We note that in NICMOS images there is no star forming region detected at the position of \object{SN 2008cs} as expected due to the high extinction towards the SN (see D{\'{\i}}az-Santos et al. 2008).

\begin{figure}[t]
\epsscale{0.90}
\plotone{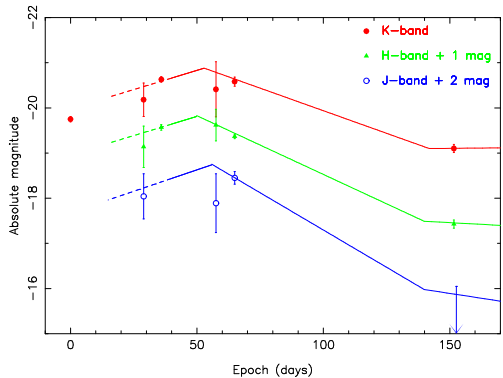}
\caption{An example fit for the SN 2008cs light curves with A$_{\rm V}(host) = 15.8$. The SN is 0.9 mag brighter than the ``slowly-declining'' CCSN template light curves and the \textit{K}-band peak occured at 53 days after the discovery, see text for details.}
\end{figure}

\subsection{The CCSN rate of IRAS 17138-1017}

To estimate the SFR for \object{IRAS 17138-1017} we compiled its 12-100 $\mu$m {\it IRAS} photometry and fitted the resulting spectral energy distribution (SED; see Fig.~3) with the starburst models of Efstathiou et al. (2000). The starburst model provides a very good fit to the near- to far-IR SED and is consistent with a negligible contribution from an active galactic nucleus (AGN; see also DePoy et al. 1988). The model assumes an exponentially decaying starburst with a time constant of 30 Myr and an age of 57 Myr. The initial A$_{\rm V}$ of the molecular clouds that constitute the starburst is 100 mag. The SFR at the peak is 169 M$_{\odot}$~yr$^{-1}$, and 76 M$_{\odot}$~yr$^{-1}$ when averaged over the duration of the starburst. We estimate a bolometric luminosity of 2.3 $\times$ 10$^{11}$ L$_{\odot}$ for \object{IRAS 17138-1017} which is close to its IR luminosity as expected for a LIRG. Adopting the average SFR and assuming CCSN progenitors between 8 and 50 M$_{\odot}$~and a Salpeter initial mass function with cut-offs at 0.1 and 125M$_{\odot}$, we obtain a CCSN rate of $\sim$0.53 yr$^{-1}$ for \object{IRAS 17138-1017}. Using 76 M$_{\odot}$~yr$^{-1}$ as an upper limit for the SFR averaged over 0.5 Gyr we obtain an upper limit for its Type Ia SN rate (Sullivan et al. 2006) of $\sim$0.03 yr$^{-1}$ which is $\sim$20 times smaller than the estimated CCSN rate.

\begin{figure}[t]
\epsscale{0.90}
\plotone{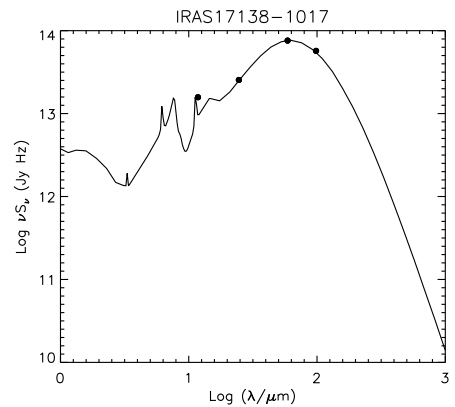}
\caption{
The IR SED of IRAS 17138-1017 fitted with the starburst models of Efstathiou
et al. (2000). No AGN contribution is needed to explain the SED.}
\end{figure}

\subsection{VLA observations}

We observed \object{IRAS 17138-1017} on 2008 May 19.4 UT (program AP558, PI: M. P\'erez-Torres) at 8.4 and 22.4 GHz, using the Very Large Array (VLA) in the C configuration, aimed at detecting any radio emission from \object{SN 2008cs} with rapid response exploratory VLA time. The observations lasted for 2.5 hr, with effective on-source times on the nucleus of \object{IRAS 17138-1017} of $\sim$60 and $\sim$49 minutes at 8.4 and 22.4 GHz, respectively.  We used the source 1713-130 for phase and amplitude calibration purposes, and the quasar 3C~286 (1331+305) to set the absolute flux density scale.  We then edited, calibrated, and imaged our 8.4 and 22.4 GHz VLA data, following standard data reduction techniques implemented within the NRAO Astronomical Image Processing System (AIPS). Since the VLA was in C-configuration, the angular resolution of the 8.4 GHz observations did not allow the detection of any compact radio emission, apart from that of the nuclear region. But our 22.4 GHz (beam size $\sim$ 0.9\arcsec) observations clearly showed a local maximum of radio emission ($S_\nu = 440 \pm 75 \mu$Jy, P\'erez-Torres et al. 2008) at R.A. $= 17^{\mathrm{h}}16^{\mathrm{m}}35.853^{\mathrm{s}}$, Decl. $= -10^{\circ}20\arcmin43.05\arcsec$. The radio source appears coincident within the uncertainties (0.2\arcsec\ in each coordinate) with the position of the SN at near-IR
wavelengths (offset of 0.03\arcsec\ and 0.07\arcsec\ in R.A. and Decl., respectively), which confirms that both the near-IR and radio emission come from the same region (see Fig~4). Future higher angular resolution observations should also detect SN 2008cs at longer wavelengths.

Type Ia SNe are not expected to be strong radio emitters and have not been detected at radio wavelengths (e.g., Panagia et al. 2006). Our detection of \object{SN 2008cs} at 22.4 GHz indicates a prominent interaction of the SN ejecta with the circumstellar medium, confirming that the SN was a core-collapse event. \object{SN 2008cs} is only the third SN after SNe 2000ft (Colina et al. 2001) and 2004ip (P{\'e}rez-Torres et al. 2007) detected at both optical/IR and radio wavelengths in a LIRG/ULIRG host.

\begin{figure}[t]
\epsscale{1.10}
\plotone{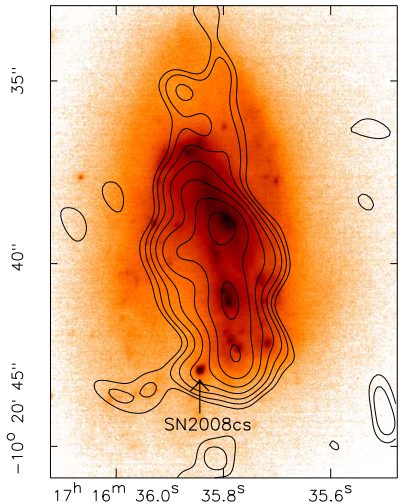}
\caption{\textit{K}-band Gemini ALTAIR/NIRI image from June 25, 2008 with 22.4 GHz VLA-C contours from May 19, 2008 overlaid.}
\end{figure}

\subsection{The ``historical'' supernova from 2004}

We carried out the photometry of the 2004 event with circular apertures and subtracted the local background. The aperture corrections were done using synthetic PSFs generated with TinyTim (Krist et al. 1998). An asteroid origin for the 2004 source was immediately ruled out since it is stationary compared to the field stars between the first and the last target frame separated by $\sim$40 minutes. Also no known minor planets were near the source within a radius of 10\arcmin\ at the time of the observation. A variable foreground star also provides a very improbable alternative explanation given the small FOV ($19.8\arcsec \times 19.8\arcsec$) of the \textit{HST}/NICMOS frames. For example, the likelihood of catching a foreground nova would be extremely small given the Galactic nova rate of $\sim$35 yr$^{-1}$ (Shafter 1997). Based on this and the high CCSN rate of \object{IRAS 17138-1017}, we conclude that the 2004 transient source was very likely a CCSN. From its (\textit{J-H}) color and assuming a negligible difference between the F160W and the ground-based \textit{H} filter and a conversion from the F110W to the ground-based \textit{J}-band photometry (Origlia \& Leitherer 2000) as well as the template CCSN light curves from Mattila \& Meikle (2001) we can estimate a host galaxy extinction for the SN. We obtain A$_{\rm V} \approx 4$ mag if it was an ``ordinary'' CCSN near the peak and A$_{\rm V} \approx 0$ mag if it was a ``slowly-declining'' CCSN at $\sim$4 months from the peak.

\section{Conclusions}

Near-IR-discovered SNe in LIRGs and ULIRGs are not only important for SN rate estimates. They can also be used to estimate the amount and distribution of extinction within their host galaxies, and follow-up observations can provide us with a better understanding of the behavior of SNe within their dusty, high-density environments. To our knowledge the host galaxy extinction of A$_{\rm V} = 15.7 \pm 0.8$ mag measured for \object{SN 2008cs} is the highest observed for a SN so far. Previous events with high extinctions include \object{SN 2001db} (A$_{\rm V} \approx 5.5$; Maiolino et al. 2002), \object{SN 2002cv} (A$_{\rm V} \approx 9$; Di Paola et al. 2002, Elias-Rosa et al. 2008), \object{SN 2002hh} (A$_{\rm V} \approx 5$; Pozzo et al. 2006), \object{SN 2004am} (A$_{\rm V} \approx 5$; Mattila et al., in preparation), and \object{SN 2004ip} (A$_{\rm V} \approx 5 - 40$; Mattila et al. 2007). Such high extinction events can have an important impact for the SN statistics when estimating the {\it complete} CCSN rates including also the optically obscured SNe. This will be essential when using CCSNe as probes of the SFR at both low- and high-z (e.g., Cappellaro et al. 1999, Dahlen et al. 2004, Botticella et al. 2008, Melinder et al. 2008) with the aim of providing a new independent measurement of the cosmic star formation history.

\acknowledgements
We thank Peter Meikle for useful comments and Harry Lehto for useful discussions. E.~K. acknowledges support from the Finnish Academy of Science and Letters (Vilho, Yrj\"{o} and Kalle V\"{a}is\"{a}l\"{a} Foundation). S.~M. and J.~K. acknowledge support from the Academy of Finland (projects 8120503 and 8121122, respectively). M.A.~P-T., A.~A. and C.~R-C. acknowledge support from the Spanish grant AYA2006-14986-C02-C01. T.~D.-S., A.~A.-H., and L.~C. acknowledge support from the grant ESP2007-65475-C02-01.

\end{document}